\documentclass{bmcart}

\usepackage[utf8]{inputenc}
\usepackage{natbib}

\usepackage[english]{babel}

\def\includegraphics{}

\startlocaldefs
\endlocaldefs

\usepackage[fleqn]{amsmath}
\usepackage{amssymb,amsfonts}
\usepackage{dsfont}                 
\usepackage{scalerel,stackengine}   
\usepackage{bm}
\usepackage{multirow}
\usepackage{url}
\usepackage{xspace}
\usepackage{csquotes}

\usepackage{tikz}
\usetikzlibrary{backgrounds}
\usetikzlibrary{arrows}
\usetikzlibrary{shapes,shapes.geometric,shapes.misc}

\pgfdeclarelayer{edgelayer}
\pgfdeclarelayer{nodelayer}
\pgfsetlayers{background,edgelayer,nodelayer,main}

\tikzstyle{none}=[inner sep=0mm]

\tikzstyle{every loop}=[]

\newcommand*{\tran}{{\mkern-1.5mu\mathsf{T}}}

\stackMath
\newcommand\reallywidehat[1]{%
\savestack{\tmpbox}{\stretchto{%
  \scaleto{%
    \scalerel*[\widthof{\ensuremath{#1}}]{\kern.1pt\mathchar"0362\kern.1pt}%
    {\rule{0ex}{\textheight}}
  }{\textheight}%
}{2.4ex}}%
\stackon[-6.9pt]{#1}{\tmpbox}%
}
\newcommand{\AUC}{\text{AUC}_e}

\def\oplogit{\mathop{\sf logit}}
\newcommand{\logit}[1]{\oplogit\left(#1\right)}
\renewcommand{\ln}{\mathop{\sf ln}}
\def\var{\mathop{\sf var}}

\def\ci{\mathop{\sf ci}}

\def\evar{\reallywidehat{\var}}
\newcommand{\ROC}{\operatorname{ROC}}
\newcommand{\ROCGLM}{{\operatorname{ROC}_g}}
\newcommand{\AUCdist}{\widetilde{\AUC}_{\text{ROC-GLM}}}

\newcommand{\Aone}{\textbf{A1}\xspace}
\newcommand{\Atwo}{\textbf{A2}\xspace}

\renewcommand{\tilde}{\widetilde}

\newcommand{\xb}{\bm{x}}
\newcommand{\data}{\mathcal{D}}
\newcommand{\dataset}{\{(\xb_1, y_1), \dots, (\xb_n, y_n) \}}
\newcommand{\R}{\mathbb{R}}
\newcommand{\indicator}[2]{\mathds{1}_{#1}(#2)}
\newcommand{\pos}{1}
\renewcommand{\neg}{0}

\newcommand{\survp}{S_\pos}
\newcommand{\survn}{S_\neg}

\newcommand{\esurvp}{\hat{S}_\pos}
\newcommand{\esurvn}{\hat{S}_\neg}

\newcommand{\tesurvp}{\tilde{S}_\pos}
\newcommand{\tesurvn}{\tilde{S}_\neg}

\newcommand{\fdim}{p}
\newcommand{\fh}{\hat{f}}

\newcommand{\xpi}{\xb_{\pos,i}}
\newcommand{\xni}{\xb_{\neg,i}}
\newcommand{\np}{n_\pos}
\newcommand{\nn}{n_\neg}
\newcommand{\tpr}{\mathop{\sf TPR}}
\newcommand{\fpr}{\mathop{\sf FPR}}

\newcommand{\sset}{\mathcal{F}}
\newcommand{\ssetp}{\sset_\pos}
\newcommand{\ssetn}{\sset_\neg}
\newcommand{\ssetpi}{\sset_{\pos, i}}
\newcommand{\ssetni}{\sset_{\neg, i}}

\newcommand{\ssetpk}[1]{\sset_\pos^{(#1)}}

\newcommand{\tssetp}{\tilde{\sset}_\pos}

\newcommand{\tssetpk}[1]{\tilde{\sset}_\pos^{(#1)}}

\newcommand{\noise}{\bm{r}}
\newcommand{\noisek}[1][k]{\bm{r}^{(#1)}}
\newcommand{\ltwosens}{\Delta_2(\fh)}

\renewcommand{\AUC}{AUC}

\newcommand{\ecdfp}{\hat{F}_\pos}
\newcommand{\ecdfn}{\hat{F}_\neg}

\newcommand{\pvsetp}{\mathcal{S}_\pos}
\newcommand{\pvsetn}{\mathcal{S}_\neg}

\newcommand{\datak}[1]{\data^{(#1)}}
\newcommand{\nk}[1][k]{n^{(#1)}}

\newcommand{\datarocglm}{\data_{\text{ROC-GLM}}}

\newcommand{\datarocglmk}[1]{\data_{\text{ROC-GLM}}^{(#1)}}

\renewcommand{\mathbf}{\bm}
\newcommand{\AUCdiff}{\AUC - \AUCdist}

\newcommand{\site}{site\xspace}
\newcommand{\sites}{sites\xspace}

\renewcommand{\th}[1][th]{\vphantom{x}^{\text{#1}}}

\begin{document}

\begin{frontmatter}

\begin{fmbox}

\title{Distributed non-disclosive validation of predictive models by a modified ROC-GLM}

\author[
  addressref={aff1,aff3,aff4},                   
  corref={aff1},                       
  email={daniel.schalk@stat.uni-muenchen.de}   
]{\inits{D.S.}\fnm{Daniel} \snm{Schalk}}
\author[
  addressref={aff2,aff3},
  email={}
]{\inits{V.S.H.}\fnm{Verena S.} \snm{Hoffmann}}
\author[
  addressref={aff1,aff4},
  email={}
]{\inits{B.B.}\fnm{Bernd} \snm{Bischl}}
\author[
  addressref={aff1, aff2,aff3},
  email={}
]{\inits{U.M}\fnm{Ulrich} \snm{Mansmann}}


\address[id=aff1]{
  \orgdiv{Department of Statistics},             
  \orgname{LMU Munich},          
  \city{Munich},                              
  \cny{Germany}                                    
}
\address[id=aff2]{
  \orgdiv{Institute for Medical Information Processing, Biometry and Epidemiology},             
  \orgname{LMU Munich},          
  \city{Munich},                              
  \cny{Germany}                                    
}
\address[id=aff3]{
  \orgdiv{DIFUTURE (DataIntegration for Future Medicine, www.difuture.de)},             
  \orgname{LMU Munich},          
  \city{Munich},                              
  \cny{Germany}                                    
}
\address[id=aff4]{%
  \orgdiv{Munich Center for Machine Learning (MCML)}
}

\end{fmbox}


\begin{abstractbox}

\begin{abstract} 
\parttitle{Background} 
Distributed statistical analyses provide a promising approach for privacy protection when analyzing data distributed over several databases. This approach brings the analysis to the data, rather than the data to the analysis. Instead of directly operating on data, the analyst receives anonymous summary statistics, which are combined into an aggregated result. Further, in model development, it is key to evaluate a trained model w.r.t.~to its prognostic or predictive performance. For binary classification, one technique is analyzing the receiver operating characteristics (ROC). Hence, we are interested to calculate the area under the curve (AUC) and ROC curve for a binary classification task using a distributed and privacy-preserving approach.

\parttitle{Methods} 
We employ DataSHIELD as the technology to carry out distributed analyses, and we use a newly developed algorithm to validate the prediction score by conducting distributed and privacy-preserving ROC analysis. Calibration curves are constructed from mean values over sites. The determination of ROC and its AUC is based on a generalized linear model (GLM) approximation of the ROC curve, the ROC-GLM, as well as on ideas of differential privacy (DP). DP adds noise (quantified by the $\ell_2$ sensitivity $\ltwosens$) to the data. The appropriate choice of the $\ell_2$ sensitivity was studied by simulations.

\parttitle{Results} 
In our simulation scenario, the true and distributed AUC measures differ by $\Delta \text{AUC} < 0.01$ depending on the choice of the differential privacy parameters. It is recommended to check the accuracy of the distributed AUC estimator in specific simulation scenarios when $\ltwosens > 0.07$. Here, the accuracy of the distributed AUC estimator may be impaired by too much artificial noise added from DP.

\parttitle{Conclusions} 
The applicability of our algorithms depends on the sensitivity of the underlying statistical/predictive model. The simulations carried out have shown that the approximation error is acceptable for the majority of simulated cases. For models with high sensitivity, the privacy parameters must be set accordingly higher to ensure sufficient privacy protection, which affects the approximation error. This work shows that complex measures, as the AUC, are applicable for validation in distributed setups while preserving an individual's privacy.

\end{abstract}

\begin{keyword}
\kwd{Area under the ROC curve}
\kwd{Distributed computing}
\kwd{Medical tests}
\kwd{ROC-GLM}
\end{keyword}


\end{abstractbox}
%

\end{frontmatter}

\section{Introduction}\label{sec:intro}

Medical research is based on the trust that the analysis of confidential patient data follows principles of privacy protection. However, depending on the released data and proposed diagnosis, breaches of the patient's privacy may occur~\citep{loukides2010disclosure}. Even when a patient gives informed consent that the researcher can have access to their pseudonymized patient data, it is necessary to keep data in a protected environment and to process it accordingly. As described by~\citet{arellano2018privacy}, when the protection of sensitive patient data is a key objective, privacy-preserving modelling should be considered. Typically, multi-centre studies in medicine or epidemiology collect the data in a central study database and perform the analyses in a specifically protected environment following the informed consent of the study subjects. Analogously, in big-data real-world applications, the data of interest may be provided by different locations and may be transferred from there to a central database for analysis. However, this requires an administratively challenging and time-consuming trustworthy data-sharing process.

Using only anonymous and aggregated data for analysis can alleviate the administrative load for data sharing. By reducing the administrative work of conventional data sharing, the new concept of employing distributed data networks in clinical studies makes it possible to leverage routinely collected electronic health data and thus streamline data collection. Non-disclosing distributed analysis is an important part of this concept, as this approach enables statistical analyses without sharing individual patient data (IPD) between the various \sites of a clinical study or sharing IPD with a central analysis unit. Thus, non-disclosing distributed analyses protect patient data privacy and enhance data security, making this a potentially advantageous approach for medical research involving sensitive patient data. However, innovative methods are needed to support robust multivariable-adjusted statistical analysis without the need to centralize IPD, thereby providing better protection for patient privacy and confidentiality in multi-database studies.

As a part of the German Medical Informatics Initiative\footnote{\url{www.medizininformatik-initiative.de}} (MII) the Data Integration for Future Medicine (DIFUTURE) consortium~\citep{DIFUTURE2018} undertakes distributed data network studies and provides tools as well as algorithms for non-disclosing distributed analyses. DIFUTURE's specific objective is to provide digital tools for individual treatment decisions and prognosis. Therefore, the development of distributed algorithms for the discovery and validation of prognostic and predictive rules is highly relevant for this mission. In the following paper, we investigate how the area under the curve (AUC) confidence intervals (CIs) proposed by \citet{delong1988comparing} behave if the computed AUC uses a generalized linear model (GLM) approach of \citet{Pepe2003} in a distributed differential privacy framework.

The concept of differential privacy was operationalized by \citet{dwork2006differential}. An algorithm is considered to be differentially private if an observer cannot determine based solely on the output whether a particular individual's information was used in the computation. Differential privacy is often discussed in the context of ensuring protection of patient data privacy, as differentially private algorithms are more likely to resist identification and reidentification attacks~\citep{dwork2006calibrating} than alternative approaches.


One of the state-of-the-art of prognostic/predictive validation techniques in a binary classification setting is to calculate the receiver operator characteristic (ROC) curve and its AUC in the pooled data as well as assess the quality of calibration~\citep{van2019calibration}. In general, IPD transfer requires specific patient consent, and data protection laws apply. Here, we present a non-disclosing distributed ROC-GLM, which we use to calculate the ROC curve, its AUC, and the respective CIs. These methods and their implementation in DataSHIELD framework~\citep{gaye2014datashield} allow analyses in which IPD does not leave its secured environment. This way, only noisy IPD under differential privacy or anonymous and aggregated statistics are shared, thereby preventing the identification of individuals. We also demonstrate that assessing the calibration of binary classification rules is a straightforward task. Thus, non-disclosing distributed validation of prediction and prognostic tools is a milestone in data-driven medicine and can unlock a plethora of medical information for research.


\paragraph{Contribution}

The work herein provides new privacy-preserving algorithms adapted to the distributed data setting for the ROC-GLM ~\citep{pepe2000interpretation}, the AUC derived therefrom, and its CIs for that AUC. To validate the algorithms, we provide a simulation study to assess estimation accuracy and to compare the results to the standard approach. Furthermore, we apply the proposed algorithms to validate a given prognostic rule on data of breast cancer patients.

We describe how the adjustments of the distributed ROC analysis are incorporated into (1) the ROC-GLM by using differential privacy~\citep{dwork2006differential} to obtain a privacy-preserving survivor function that can be communicated without the threat of privacy breaches and (2) secure aggregations to conduct a distributed Fisher scoring algorithm~\citep{jones2013combined} to obtain parameter estimates for the ROC-GLM. In addition to the ROC analysis to assess the discrimination of a classifier, we describe a distributed calibration approach that respects the privacy of the individuals. Furthermore, we introduce a distributed version of the Brier score~\citep{brier1950verification} and the calibration curve~\citep{vuk2006roc}.

\section{Related literature}

\citet{boyd2015differential} calculate the AUC under differential privacy using a symmetric binormal ROC function. However, our approach is more general, with a possible extension to multiple covariates. While they derive the AUC from the ROC parameters, we use integration techniques. We also provide CIs for the AUC. \citet{unal2021ppaurora} use homomorphic encryption to calculate the ROC curve. Their approach does not provide CIs or an extension to multiple covariates. To the best of our knowledge, a modified ROC-GLM algorithm for non-disclosing distributed analyses has so far not been developed.


\section{Background}\label{sec:terms}

Throughout this paper, we consider binary classification, with $\pos$ for a case with the trait(s) of interest (i.e., \enquote{diseased}, \enquote{success}, \enquote{favorable}) and $\neg$ for the remaining cases (i.e., lacking trait(s) of interest, \enquote{healthy}, \enquote{no success}, \enquote{unfavorable}). Furthermore, $f(\xb)\in\R$ is the true score based on a true but unknown function $f$ for a patient with a feature vector $\xb$ of an underlying random vector $\bm{X}$. In this paper, this score can also express a posterior probability with $f(\xb) \in [0,1]$ and is explicitly noted in the corresponding text passages. The function $f$ is estimated by a statistical (classification) model $\fh:\R^\fdim \rightarrow \R$. The estimated individual score for a subject with feature or covariate vector $\xb\in\R^\fdim$ is $\fh(\xb)$. The training data set used to fit $\fh$ is denoted as $\data = \dataset$ with $y_i \in \{\pos, \neg\}$. The score $\fh(\xb)$ and a threshold value $c\in\R$ are used to build a binary classifier:  $\indicator{[c,\infty)}{\fh(\xb)}$. On an observational level, $\xpi$ and $\xni$ indicate the $i\th$ observation that corresponds to a positive or negative output $y$. The number of observations in $\data$ with output $\pos$ and $\neg$ are denoted by $\np$ and $\nn$. The set of scores that corresponds to the positive or negative output is denoted by $\ssetp = \{\fh(\xpi)\ |\ i = 1, \dots, \np\}$ and $\ssetn = \{\fh(\xni)\ |\ i = 1, \dots, \nn\}$, with $\ssetpi = \fh(\xpi)$ and $\ssetni = \fh(\xni)$.

\subsection{ROC curve and AUC}

To quantify the quality of a binary classifier, we use the true positive rate (TPR) and false positive rate (FPR) with values between $0$ and $1$: $\tpr(c) = P(f(\bm{X}) \geq c\ |\ Y = 1)$ and $\fpr(c) = P(f(\bm{X}) \geq c\ |\ Y = 0)$ for threshold $c\in\R$~\citep{pepe2000interpretation}. These probability functions are also known as \textit{survivor functions} $\survp(c) = \tpr(c)$ and $\survn(c) = \fpr(c)$. The ROC curve is defined as $\ROC(t) = \survp(\survn^{-1}(t))$. The AUC as a measure of discrimination between the two distributions of the positive and negative class is given as $\AUC = \int_0^1 \ROC(t)\ dt$~\citep{zweig1993receiver}.

\subsection{Empirical calculation of the ROC curve and AUC}\label{subsec:emp-auc-and-pv}

The calculation of the empirical ROC curve uses the \textit{empirical survivor functions} $\esurvp$ and $\esurvn$. These functions are based on the empirical cumulative distribution functions (ECDF) $\ecdfp$ of $\ssetp$ and $\ecdfn$ of $\ssetn$: $\esurvp = 1 - \ecdfp$ and $\esurvn = 1 - \ecdfn$. The set of possible values of the empirical TPR and FPR are given by $\pvsetp = \{\esurvp(\fh(\xni))\ |\ i = 1, \dots, \nn\}$ and $\pvsetn = \{\esurvn(\fh(\xpi))\ |\ i = 1, \dots, \np\}$ and are also called \textit{placement values}. These values standardize a given score relative to the class distribution~\citep{Pepe2003}. The set $\pvsetp$ represents the positive placement values and $\pvsetn$ the negative placement values.

The empirical version of the $\ROC(t)$ is a discrete function derived from the placement values $\pvsetp\subseteq \{0, 1/\np, \dots, (\np-1)/\np, 1\}$ and $\pvsetn\subseteq\{0, 1/\nn, \dots, (\nn-1)/\nn, 1\}$. The empirical AUC is then a sum over rectangles of width $1 / \nn$ and height $\esurvp(\fh(\xni))$:
\begin{equation}
    \widehat{\AUC} = \nn^{-1}\sum_{i=1}^{\nn} \esurvp(\fh(\xni)).
\end{equation}

\subsection{CI for the empirical AUC}\label{subsec:emp-auc-ci}

The approach proposed by~\cite{delong1988comparing} is used to calculate CIs. Here, the variability of the estimated AUC from the empirical variance ($\evar$) of the placement values is determined by:
\begin{equation}\label{eq:auc-var}
    \evar(\AUC) = \frac{\evar(\pvsetp)}{\nn} + \frac{\evar(\pvsetn)}{\np}.
\end{equation}
This approach provides a CI for the logit AUC, from which the CI for the AUC can be derived by the $\oplogit^{-1}$ transformation:
\begin{equation}\label{eq:auc-ci}
    {\ci}_\alpha\left(\logit{\AUC}\right) = \logit{\widehat{\AUC}} \pm \Phi^{-1}\left(1 - \frac{\alpha}{2}\right) \frac{\sqrt{\evar\left(\AUC\right)}}{\widehat{\AUC}\left(1 - \widehat{\AUC}\right)}.
\end{equation}
The term $\Phi^{-1}$ denotes the quantile function of the standard normal distribution. Furthermore, statistical testing can be conducted based on that CI. For example, the hypothesis $H_0: \AUC \leq a_0$ vs. $H_1: \AUC > a_0$ with a significance level of $\alpha$ can be tested by checking whether $\logit{a_0} > a$, $\forall a \in \ci_\alpha$ to reject $H_0$.

\subsection{The ROC-GLM}\label{subsec:ROC-GLM}

The ROC-GLM interprets the ROC curve as a GLM \citep[Section~5.5.2]{Pepe2003}: $\ROCGLM_{}(t | \bm{\gamma}) = g(\mathbf{\gamma} h(t))$, with link function $g:\R\to [0,1], \eta\mapsto g(\eta)$, coefficient vector $\mathbf{\gamma}\in\R^l$, and covariate vector $h: \R\rightarrow\R^l, t \mapsto \bm{h}(t) = (h_1(t), \dots, h_l(t))^\tran$. The ROC-GLM is an unbiased estimator of the ROC~\citep{pepe2000interpretation}. Estimating the ROC-GLM is based on an intermediate data set $\datarocglm = \{(u_{ij}, \bm{h}(t_j))\ |\ i = 1, \dots, \np, j = 1, \dots, n_T\}$ with covariates $\bm{h}(t_j)$, set of thresholds $T = \{t_1, \dots, t_{n_T}\}$, and binary response $u_{ij}\in\{0,1\}$, $u_{ij} = \indicator{(\esurvn(\ssetpi),\infty)}{t_j} = \indicator{(-\infty,\ssetpi]}{\esurvn^{-1}(t_j)}$. The simplest ROC-GLM uses the two-dimensional vector $\bm{h}(t)$ with $h_1(t) = 1$ and $h_2(t) = \Phi^{-1}(t)$. Setting the link function to $g = \Phi$ results in the binormal form $\ROCGLM_{}(t|\bm{\gamma}) = \Phi(\gamma_1 + \gamma_2\Phi^{-1}(t))$ and is represented as a probit regression with response variable $u_{ij}$ and covariate $\Phi^{-1}(t_j)$. A common strategy for choosing the set of thresholds $T$ is to use an equidistant grid.

The estimated ROC curve $\ROCGLM(t | \hat{\bm{\gamma}})$ results from estimating the model parameters $\mathbf{\gamma}$ as $\hat{\mathbf{\gamma}}$. The estimated AUC from the ROC-GLM $\widehat{\AUC}_{\text{ROC-GLM}}$ is obtained by calculating the integral $\widehat{\AUC}_{\text{ROC-GLM}} = \int_0^1 \ROCGLM(t|\hat{\bm{\gamma}})\ dt$. Here, we use \texttt{R}s \texttt{integrate} function~\citep{piessens2012quadpack}. Figure~\ref{fig:roc-glm-single} visualizes the ROC-GLM algorithm with all individual parts.

\begin{figure}[ht]
    \centering
\tikzstyle{white text}=[fill=none, draw=none, shape=circle, text=white]

\tikzstyle{field}=[-, fill={rgb,255: red,207; green,226; blue,243}, text=white, draw=none]
\tikzstyle{arrow}=[->, line width=1.5pt, fill=none, draw={rgb,255: red,11; green,83; blue,148}]
\tikzstyle{field dark}=[-, fill={rgb,255: red,11; green,83; blue,148}, draw=none]
\tikzstyle{dashed blue arrow}=[draw={rgb,255: red,11; green,83; blue,148}, dashed, line width=1pt, ->]
\tikzstyle{field final}=[-, fill={rgb,255: red,217; green,234; blue,211}, draw=none]
\tikzstyle{new edge style 0}=[-, draw={rgb,255: red,11; green,83; blue,148}]
\tikzstyle{blue line}=[-, draw={rgb,255: red,11; green,83; blue,148}, line width=1.5pt]

\begin{tikzpicture}[scale=0.4, transform shape]
    \tikzstyle{every node}=[font=\huge]
	\begin{pgfonlayer}{nodelayer}
		\node [style=none] (0) at (-11, 7.5) {};
		\node [style=none] (1) at (-11, 5.5) {};
		\node [style=none] (2) at (-7, 5.5) {};
		\node [style=none] (3) at (-7, 7.5) {};
		\node [style=none] (5) at (-9, 6.5) {Model $\hat{f}$};
		\node [style=none] (6) at (-11, 2.5) {};
		\node [style=none] (7) at (-11, 0.5) {};
		\node [style=none] (8) at (-7, 0.5) {};
		\node [style=none] (9) at (-7, 2.5) {};
		\node [style=none] (10) at (-9, 1.5) {};
		\node [style=none] (11) at (-9, 1.5) {Data $\mathcal{D}$};
		\node [style=none] (13) at (-7, 4) {};
		\node [style=none] (14) at (-11, 4) {};
		\node [style=none] (15) at (-9, 5.5) {};
		\node [style=none] (17) at (-9, 2.5) {};
		\node [style=none] (18) at (-9, 3.75) {};
		\node [style=none] (19) at (-9, 4.25) {};
		\node [style=white text] (22) at (-6, 4) {$\mathcal{F}$};
		\node [style=none] (23) at (-5, 4) {};
		\node [style=none] (24) at (-4.5, 4) {};
		\node [style=white text] (26) at (-3.5, 4) {$\hat{S}_0$};
		\node [style=white text] (27) at (0.5, 4) {$\mathcal{D}_{\text{ROC-GLM}}$};
		\node [style=none] (28) at (-6.75, 4.5) {};
		\node [style=none] (29) at (-6.75, 3.5) {};
		\node [style=none] (30) at (-5.25, 3.5) {};
		\node [style=none] (31) at (-5.25, 4.5) {};
		\node [style=none] (32) at (-4.25, 4.5) {};
		\node [style=none] (33) at (-4.25, 3.5) {};
		\node [style=none] (34) at (-2.75, 3.5) {};
		\node [style=none] (35) at (-2.75, 4.5) {};
		\node [style=none] (36) at (-1.5, 5) {};
		\node [style=none] (37) at (-1.5, 3) {};
		\node [style=none] (38) at (2.5, 3) {};
		\node [style=none] (39) at (2.5, 5) {};
		\node [style=none] (40) at (-2.5, 4) {};
		\node [style=none] (41) at (-2, 4) {};
		\node [style=none] (47) at (0.5, 6.5) {Thresholds $T$};
		\node [style=none] (48) at (0.5, 6) {};
		\node [style=none] (49) at (0.5, 5) {};
		\node [style=none] (50) at (-1.5, 7) {};
		\node [style=none] (51) at (-1.5, 6) {};
		\node [style=none] (52) at (2.5, 6) {};
		\node [style=none] (53) at (2.5, 7) {};
		\node [style=none] (54) at (-6.5, 1.5) {};
		\node [style=none] (55) at (0.5, 3) {};
		\node [style=none] (56) at (5.5, 5.5) {};
		\node [style=none] (57) at (5.5, 2.5) {};
		\node [style=none] (58) at (5.5, 4) {$\ROC_g$};
		\node [style=none] (59) at (3, 4) {};
		\node [style=none] (60) at (4, 4) {};
	\end{pgfonlayer}
	\begin{pgfonlayer}{edgelayer}
		\draw [style=field] (2.center)
			 to [bend right=90, looseness=0.75] (3.center)
			 to (0.center)
			 to [bend right=90, looseness=0.75] (1.center)
			 to cycle;
		\draw [style=field] (8.center)
			 to [bend right=90, looseness=0.75] (9.center)
			 to [bend right, looseness=0.00] (6.center)
			 to [bend right=90, looseness=0.75] (7.center)
			 to cycle;
		\draw [style=arrow] (14.center) to (13.center);
		\draw [style=arrow] (15.center) to (19.center);
		\draw [style=arrow] (17.center) to (18.center);
		\draw [style=arrow] (23.center) to (24.center);
		\draw [style=field dark] (28.center)
			 to (31.center)
			 to [bend left=90, looseness=0.75] (30.center)
			 to (29.center)
			 to [bend right=270, looseness=0.75] cycle;
		\draw [style=field dark] (32.center)
			 to (35.center)
			 to [bend left=90, looseness=0.75] (34.center)
			 to (33.center)
			 to [bend right=270, looseness=0.75] cycle;
		\draw [style=field dark] (39.center)
			 to [bend left=90, looseness=0.75] (38.center)
			 to (37.center)
			 to [bend left=90, looseness=0.75] (36.center)
			 to cycle
			 to (36.center);
		\draw [style=arrow] (40.center) to (41.center);
		\draw [style=arrow] (48.center) to (49.center);
		\draw [style=field] (52.center)
			 to (51.center)
			 to [bend right=270, looseness=0.75] (50.center)
			 to (53.center)
			 to [bend left=90, looseness=0.75] cycle;
		\draw [style=dashed blue arrow, in=-90, out=-15, looseness=0.75] (54.center) to (55.center);
		\draw [style=field final] (56.center)
			 to [in=0, out=0, looseness=1.75] (57.center)
			 to [bend right=270, looseness=1.75] cycle;
		\draw [style=arrow] (59.center) to (60.center);
	\end{pgfonlayer}
\end{tikzpicture}
    \caption{All parts of the $\operatorname{ROCGLM}(\data)$ procedure, starting with the data and a model for predicting scores $Y$, calculating the survivor function $\hat{S}_{\bar{D}}$, and finally calculating intermediate data $\datarocglm$ for the probit regression to obtain the parameters of the ROC-GLM $\ROCGLM$.}
    \label{fig:roc-glm-single}
\end{figure}

\subsection{Differential privacy}\label{subsec:diff-privacy}

Following \citet{dwork2014algorithmic}, we add normally distributed noise $\noise$ to a randomized mechanism $\mathcal{M}: \mathcal{X} \mapsto \mathcal{Y}$ with domain $\mathcal{X}$ (e.g., $\mathcal{X} = \R^p$) and target domain $\mathcal{Y}$ (e.g., $\mathcal{Y} = \R$ in regression) to ensure $(\varepsilon, \delta)$-differential privacy~\citep{dwork2006our}. $(\varepsilon, \delta)$-differential privacy is given if, for any subset of outputs $R \subseteq \mathcal{Y}$, the property $P(\mathcal{M}(\xb) \in R) \leq \exp(\varepsilon) P(\mathcal{M}(\xb^\prime) \in R) + \delta$ holds for two adjacent inputs\footnote{\footnotesize{In theory, multiple ways exist to define adjacent inputs. Throughout this article, adjacent inputs are based on a histogram representation $\tilde{\xb}\in\mathbb{N}^p$ and $\tilde{\xb}^\prime\in\mathbb{N}^p$ of two input vectors $\xb$ and $\xb^\prime$. The definition of adjacent inputs is then given by an equal $\ell_1$ norm of $\tilde{\xb}$ and $\tilde{\xb}^\prime$ to one: $\text{adjacent}\ \xb,\ \xb^\prime\ \Leftrightarrow \ \| \tilde{\xb} - \tilde{\xb}^\prime \|_1 = 1$ \citep[cf.,][]{dwork2014algorithmic}.}} $\xb, \xb^\prime \in \mathcal{X}$. The value of $\varepsilon$ controls how much privacy is guaranteed. The value of $\delta$ is the probability that $(\varepsilon,0)$-differential privacy is broken (also known as $\varepsilon$-differential privacy and the original definition proposed in~\citep{dwork2006calibrating}).

Our randomized mechanism is $\mathcal{M}(\xb) = \fh(\xb) + r$. Hence, the protected values of the survivor function are $\tssetp = \{\mathcal{M}(\xb_{1,i})\ |\ i = 1, \dots, \np\}$ and not the original score values $\ssetp$. The noise $\noise$ follows a normal distribution $\mathcal{N}(0, \tau^2)$. The variance is set to any value $\tau \geq c \ltwosens / \varepsilon$ with $c^2 > 2 \ln(1.25/\delta)$, $\varepsilon \in (0,1)$, and $\ltwosens$ is the $\ell_2$-sensitivity of $\fh$ measured as $\ltwosens = \max_{\text{adjacent}\ \xb,\ \xb^\prime} \|\fh(\xb) - \fh(\xb^\prime)\|_2$. In practice, we first calculate the $\ell_2$-sensitivity of the prediction model $\hat{f}$ to determine possible values for $\varepsilon$ and $\delta$ (see Section~\ref{subsubsec:results-AUC}). Then, we control the amount of noise added to the algorithm by choosing $\varepsilon$ and $\delta$, which sets the variance of the generated noise via $\tau = c \ltwosens / \varepsilon$ and $c = \sqrt{2 \ln(1.25 / \delta)}$. Appendix~A.2. contains further details and a visualization of the Gaussian mechanism.

\section{Distributed ROC-GLM}

\newcommand{\vb}{\mathbf{v}}

\subsection{General principles}\label{subsec:gen-princ}

A total of $K$ data sets are distributed over a network of $K$ \sites: $\datak{1}, \dots, \datak{K}$. Each data set $\datak{k}$ consists of $\nk$ observations $(\xb_i^{(k)}, y_i^{(k)})$. The $i\th$ feature vector of the $k\th$ \site is denoted by $x_{.,i}^{(k)}$. The $i\th$ outcome on \site $k$ is $y_i^{(k)}$. We assume the distributed data to be part of the full but inaccessible data set:
\begin{equation}\label{eq:distr-data}
    \data =\bigcup_{k=1}^K \datak{k},\ \ n = \nk[1] + \dots + \nk[K]
\end{equation}

Instead of calculating the ROC-GLM for one local data set, we want to calculate the ROC-GLM on $K$ confidential distributed data sets $\datak{1}, \dots, \datak{K}$. All shared information must comply with the following non-disclosing principles:
\begin{itemize}
    \item[\Aone]
        Aggregated values from which it is not possible to derive original values are shared. Therefore, an aggregation $a: \R^d \mapsto \R$, $\vb \rightarrow a(\vb)$ with $d \geq q \in\mathbb{N}$ must be applied to allow sharing the value $a(\vb)$. The value of $q$ is a \textit{privacy level} guaranteeing that at least $q$ values were used to gain $a(\vb)$. In the distributed setup, the aggregation $a(\vb^{(k)})$ with $\nk[k]$ unique values in $\vb^{(k)}$ shared from each of the $K$ \sites can then be further processed. Values $a(\vb^{(k)})$ are just allowed to be shared if $\nk[k] \geq q$.


    \item[\Atwo]
        Differential privacy~\citep{dwork2006differential} is used to ensure non-disclosive IPD via a noisy representation.
\end{itemize}

\paragraph{Example: Distributed Brier score and calibration curve}

Probabilistic (or scoring) classifiers can be assessed by quantifying discrimination and calibration. While the AUC measures discrimination, calibration is often addressed by the Brier score~\citep{brier1950verification} or a calibration curve~\citep{vuk2006roc}. Both can be calculated by considering criterion $\Aone$.

\textit{Brier score:} The Brier score $\operatorname{BS}$ is defined as the mean squared error of the true 0-1-labels and the predicted probabilities of belonging to class 1. For the Brier score, the score $\fh(\xb) \in [0,1]$ is given as posterior probability. The Brier score is calculated by:
\begin{equation}
    \operatorname{BS} = n^{-1}\sum_{i=1}^n \left(y_i - \fh(\xb_i)\right)^2
\end{equation}
Hence, having a prediction model $\hat{f}$ at each of the $K$ \sites, we can calculate the Brier score by:
\begin{enumerate}
    \item
        Calculating the residuals $e_i^{(k)}$ based on the true label $y_i^{(k)}$ at \site $k$ and the predicted probabilities $\hat{f}(\xb_i^{(k)})$: $e_i^{(k)} = y_i^{(k)} - \hat{f}(\xb_i^{(k)})$, $\forall i = 1, \dots, n^{(k)}$.

    \item
        Calculating $a_{\text{sum}}(\bm{e}^{(k)} \circ \bm{e}^{(k)})$, with $\bm{e}^{(k)} = (e_1^{(k)}, \dots, e_{\nk[k]}^{(k)})^\tran\in\R^{n_k}$, the element-wise product $\circ$, and aggregation $a_{\text{sum}}(\vb^{(k)}) = \sum_{i=1}^{\nk[k]} v_i^{(k)}$.

    \item
        Sending $a_{\text{sum}}(\bm{e}^{(k)} \circ \bm{e}^{(k)})$ and $\nk[k]$ (if $n_k \geq q$) to the host, who finally calculates $\operatorname{BS} = n^{-1}\sum_{k=1}^K a_{\text{sum}}(\bm{e}^{(k)} \circ \bm{e}^{(k)})$.
\end{enumerate}

\textit{Calibration curve:} To calculate a calibration curve, we discretize the domain of the probabilistic classifier $\hat{f}$ in $[0,1]$ into $n_{\text{bin}}$ bins (for example, $n_{\text{bin}} + 1$ equidistant points $p_i$ from $0$ to $1$ to construct the $n_{\text{bin}}$ bins $b_l = [p_l, p_{l+1})$ for $l = 1, \dots, n_{\text{bin}} - 1$ and $b_{n_{\text{bin}}} = [p_{n_{\text{bin}}}, p_{n_{\text{bin}} + 1}]$ for $l = n_{\text{bin}}$). The calibration curve is the set of $2$-dimensional points $p_{\text{cal},l} = (\operatorname{pf}_l, \operatorname{tf}_l)$, with $\operatorname{tf}_l = |\mathcal{I}_l|^{-1}\sum_{i \in \mathcal{I}_l} y_i$ as the true fraction of $y_i = 1$ in bin $b_l$ and $\operatorname{pf}_l = |\mathcal{I}_l|^{-1}\sum_{i \mathcal{I}_l} \hat{f}(\xb_j)$ as the predicted fraction for outcome $1$ in $b_l$. The set $\mathcal{I}_l$ describes the observations for which the prediction $\hat{f}(\xb_i)$ falls into bin $b_l$: $\mathcal{I}_l = \{i \in \{1, \dots, n\}\ |\ \hat{f}(\xb_i)\in b_l\}$. A probabilistic classifier $\hat{f}$ is well-calibrated if the points $p_{\text{cal},l}$ are close to the bisector.

In the distributed setup, the points $p_{\text{cal},l}$ are constructed by applying the distributed mean to both points for each bin at each \site:
\begin{enumerate}
    \item
        Set all $b_1, \dots, b_{n_{\text{bin}}}$, and communicate them to the \sites.

    \item
        Calculate the values $c_{l, \operatorname{pf}}^{(k)} = a_{\text{sum}}(\{\hat{f}(\xb_i^{(k)})\ |\ i \in \mathcal{I}_l^{(k)}\})$ and $c_{l, \operatorname{tf}}^{(k)} = a_{\text{sum}}(\{y_i^{(k)}\ |\ i \in \mathcal{I}_l^{(k)}\})$ for all $l = 1, \dots, n_{\text{bin}}$.

    \item
        Send $\{(c_{l, \operatorname{tf}}^{(k)}, c_{l, \operatorname{pf}}^{(k)}, |\mathcal{I}^{(k)}_l|)\ |\ k = 1, \dots, K, l = 1, \dots, n_{\text{bin}}\}$ to the host if $|\mathcal{I}_l^{(k)}| \geq q$. 

    \item
        The host calculates the calibration curve $p_{\text{cal},l}$ by aggregating the elements $\operatorname{tf}_l = (\sum_{k=1}^K |\mathcal{I}_l^{(k)}|)^{-1}\sum_{k=1}^K c_{l,\operatorname{tf}}^{(k)}$ and  $\operatorname{pf}_l = (\sum_{k=1}^K |\mathcal{I}_l^{(k)}|)^{-1}\sum_{k=1}^K c_{l,\operatorname{pf}}^{(k)}$ for $l = 1,\dots,n_{\text{bin}}$.
\end{enumerate}

\paragraph{Parts of the distributed ROC-GLM}

Two aspects are important to construct the distributed version of the ROC-GLM: $\operatorname{distrROCGLM}$. First, the distributed version of the empirical survivor function; Second, a distributed version of the probit regression. Figure~\ref{fig:roc-glm-parallel} shows details of the general procedure. The starting point of the distributed ROC-GLM is the private data $\datak{1}, \dots, \datak{K}$ on the $K$ \sites.

The global survivor function $\esurvn$ is approximated by $\tesurvn$ (Section~\ref{subsec:distr-survivor-fct}) using principle \Atwo. The computation of $\tesurvn$ depends on the level of privacy induced by the $(\varepsilon,\delta)$-differential privacy parameters (Section~\ref{subsec:diff-privacy}). The accuracy of the AUC as well as its CI depends on the choice of $\varepsilon$ and $\delta$. The global survivor function $\tesurvn$ is transmitted to each of the $K$ \sites and allows calculation of a local version of the intermediate data set $\datarocglmk{k}$ (See Section~\ref{subsec:ROC-GLM}). The distributed probit regression complies with principle \Aone and produces the distributed ROC-GLM parameter estimates (see Section~\ref{subsec:distr-glm}). Using the ROC-GLM of these parameters, denoted by $\widetilde{\ROC}_g$, allows calculation of the approximated AUC, denoted by $\AUCdist = \int_0^1 \widetilde{\ROC}_g(t|\hat{\bm{\gamma}})\ dt$. Finally, the CIs can be calculated based on a variance estimation, which also complies with principle \Aone (See Section~\ref{subsec:distr-ci}).
\begin{figure}[ht]
    \centering
    \input{figures/roc-glm-distr.tikz}
    \caption{All parts of the $\operatorname{distrROCGLM}$ procedure calculating the distributed approximation $\widetilde{\ROC}_g$ of $\ROCGLM$. The starting points are the \sites (here $K = 3$), which communicate scores with added noise, calculate the global survivor function $\tilde{S}_{\bar{D}}$, and finally calculate the distributed probit regression on intermediate data $\datarocglmk{k}$ at each \site.}
    \label{fig:roc-glm-parallel}
\end{figure}

\subsection{Approximating the global survivor functions}\label{subsec:distr-survivor-fct}

The greatest challenge here is the privacy-preserving calculation of the global survivor function. It is prohibited to directly communicate score values $\ssetpk{k}$ from the local \sites to the analyst. Instead, we propose to calculate an approximation $\tesurvp$: First, we set the value of $\varepsilon$ and $\tau$ and generate a noisy representation $\tssetpk{k} = \ssetpk{k} + \mathbf{r}^{(k)}$ of the original score values $\ssetpk{k}$ at each site. Second, the noisy scores are communicated to the host and pooled to $\tssetp = \bigcup_{k=1}^K \tssetpk{k}$ to calculate an approximation $\tesurvp$ of the global survivor function. Third, $(\varepsilon, \delta)$-differential privacy allows sharing $\tesurvp$ with all \sites. Forth, the local sites calculate the global placement values and create the intermediate data set to enter the distributed probit regression.

\subsection{Distributed GLM}\label{subsec:distr-glm}

Existing solutions for distributed computing -- such as federated learning~\citep{mcmahan2017communication} -- are based on an iterative process of sharing and aggregating parameter values. Although this approach could also be applied to GLMs, it may lead to inexact estimates for heterogeneous data situations~\citep{yang2021characterizing}. For distributed calculation of the GLM, we use an approach described by~\cite{jones2013combined} and adjust the optimization algorithm of GLMs -- the Fisher scoring -- at its base to estimate parameters without performance loss. This approach complies with \Aone.

The basis of the ROC-GLM is a probit regression (and therefore a GLM) with $\mathbb{E}(Y\ |\ X = x) = g(x^\tran \theta)$ with link function $g$, response variable $Y$, and covariates $X$. The Fisher scoring is an iterative descending technique $\hat{\theta}_{m+1} = \hat{\theta}_m + \mathcal{I}^{-1}(\hat{\theta}_m)\mathcal{V}(\hat{\theta}_m)$ that uses second order gradient information. The components are the score vector $\mathcal{V}(\hat{\theta}_m) = [ \partial \ell_\theta(y,x) / \partial \theta ]_{\theta = \hat{\theta}_m} \in \R^{p}$ and the observed Fisher information $\mathcal{I}(\hat{\theta}_m) = [\partial \mathcal{V}(\theta) / \partial \theta ]_{\theta = \hat{\theta}_m} \in \R^{p\times p}$ based on the log likelihood $\ell_\theta(\mathcal{D}) = \sum_{i=1}^n \log(f_Y(y_i,x_i))$. A common stop criterion (as used in \texttt{R}s~\citep{rcore} \texttt{glm} function) to determine whether the Fisher scoring has converged or not is when the relative improvement $|dev_m - dev_{m-1}| / (|dev_m| + 0.1)$ of the deviance $dev_m = -2\ln( \ell_{\hat{\theta}_m}(\mathcal{D}))$ is smaller than a value $a$. The default value used in the \texttt{glm} function of \texttt{R} is $a = 10^{-8}$.

With non-overlapping data at the $K$ \sites (each subject contributes information only at a unique \site), condition~\eqref{eq:distr-data} is fulfilled. This implies the additive structure of the global score vector $\mathcal{V}(\theta_m)$ and Fisher information $\mathcal{I}(\theta_m)$. With the \site-specific score vector $\mathcal{V}_k(\theta_m)$ and Fisher information $\mathcal{I}_k(\theta_m)$, it holds:
\begin{align}
    \mathcal{V}(\hat{\theta}_m) &= \sum\limits_{k=1}^K \mathcal{V}_k(\hat{\theta}_m) \label{eq:score-deomposition} \\
    \mathcal{I}(\hat{\theta}_m) &= \sum_{k=1}^K \mathcal{I}_k(\hat{\theta}_m) \label{eq:fisher-decomposition}
\end{align}

This process complies with \Aone and allows estimation of the parameter vector $\hat{\theta}$ with the same precision as in an analysis based on data aggregated over the \sites.

\subsection{Distributed CIs for the AUC}\label{subsec:distr-ci}

A straightforward consequence from Section~\ref{subsec:gen-princ}, is that the distributed calculation of the global sample mean ($\operatorname{distrAVG}(\vb^{(1)}, \dots, \vb^{(K)})$) complies with \Aone. Here, we provide a distributed version of the sample variance $\evar(\vb) = (n-1)^{-1} \sum_{i=1}^n (v_i - \bar{v})^2$ by a two-step procedure. In the first step, the sample mean is calculated using $\bar{v} = \operatorname{distrAVG}(\vb^{(1)}, \dots, \vb^{(K)})$ and shared with all $K$ \sites. In the second step, each \site calculates the aggregation $a_{\text{var}}(\vb^{(k)}) = \sum_{i=1}^{\nk} (v_i^{(k)} - \bar{v})^2$, which is further aggregated to the sample variance $\evar(\vb) = (n - 1)^{-1}\sum_{k=1}^K a_{\text{var}}(\vb^{(k)})$: $\operatorname{distrVAR}(\vb^{(1)}, \dots, \vb^{(K)})$. The operations $\operatorname{distrAVG}$ and $\operatorname{distrVAR}$ fulfill \Aone if $\nk \geq q$, $\forall k \in \{1, \dots, K\}$.

Based on operation $\operatorname{distrVAR}$, non-disclosing distributed CIs for the global AUC can be provided. As described in Section~\ref{subsec:emp-auc-and-pv} and Section~\ref{subsec:emp-auc-ci}, the calculation of the approximated CI requires both approximated survivor functions $\tesurvn$ and $\tesurvp$ (see Section~\ref{subsec:distr-survivor-fct}). A distributed CI $\widetilde{\ci}_\alpha$ to approximate $\ci_\alpha$ follows from Formula~\eqref{eq:auc-ci}.

\section{Simulation study}

\subsection{General considerations}\label{subsec:Gen_Cons_2}

It is emphasized in Section~\ref{subsec:distr-glm} that the survivor functions for the data at hand are needed to build placement values and to create the data set for the probit regression in order to estimate the ROC curve and its AUC. Based on the Gaussian mechanism, noise is generated to create a non-disclosing distributed survivor function. The aim of the simulation study is to understand the effect of the introduced noise (which is necessary to conduct the distributed analysis) on the accuracy when compared to the empirical AUC and the CI of \citet{delong1988comparing}. We assume that the well-studied empirical AUC~\citep{hanley1982meaning, mason2002areas} and CI are adequate estimators of the true AUC of the underlying data generating process that is already attached with a certain estimation error. Our goal is not to construct better estimates for the true AUC, but to study the difference between our distributed approach to the estimates applied to the pooled data.

In our simulation, we explore the bias introduced by our distributed approach. To assess the accuracy of our distributed approach when estimating the AUC, we measure the difference $\Delta \AUC = \AUCdiff$ of the AUC obtained by the distributed ROC-GLM $\AUCdist$ (Section~\ref{subsec:gen-princ}) and the empirical $\AUC$ (Section~\ref{subsec:emp-auc-and-pv}). Of interest is obtaining an accuracy of $|\Delta \AUC | \leq 0.01$.

For the CI, we calculate the error $\Delta \ci_\alpha$ based on the symmetric difference between $\ci_\alpha$ proposed by~\citet[][see Section~\ref{subsec:emp-auc-ci}]{delong1988comparing} and our non-disclosing distributed approach $\widetilde{\ci}_\alpha$ (Section~\ref{subsec:distr-ci}). We study $\Delta \ci_\alpha = |\widetilde{\ci}_{\alpha, l} - \ci_{\alpha, l}| + |\widetilde{\ci}_{\alpha, r} - \ci_{\alpha, r}|$, with indices $l$ and $r$ denoting the left and right side of the CI, respectively. It is of interest to have an error smaller than $0.01$: $\Delta \ci_\alpha < 0.01$.

We explore the following research questions:

\newcommand{\rsone}{\textbf{Question 1}\xspace}
\newcommand{\rstwo}{\textbf{Question 2}\xspace}
\newcommand{\rsoneText}{Correctness of the ROC-GLM and distributed ROC-GLM\xspace}
\newcommand{\rstwoText}{Correctness of the AUC CIs\xspace}

\begin{itemize}
    \item[]\hspace{-2em}\rsone\textbf{-- \rsoneText} (Section~\ref{subsubsec:results-AUC}): How can we set both privacy parameters $\varepsilon$ and $\delta$ to reach $|\Delta \AUC|$ below 0.01?
    \item[]\hspace{-2em}\rstwo\textbf{-- \rstwoText} (Section~\ref{subsub:corr-cis}): How can we set both privacy parameters $\varepsilon$ and $\delta$ to reach $\Delta \ci_\alpha$ below 0.01?
\end{itemize}

\subsection{Data generation}\label{subsec:simulation-generation}

The aim of the following data generation is to simulate uniformly distributed AUC values between 0.5 and 1. (1) The data generation starts with randomly picking $n$ from $\{100, \dots, 2500\}$. (2) For each $i \in\{1, \dots, n\}$, the \textit{true} prediction scores are generated from the uniform distribution $\sset_i\sim U[0,1]$. Next, (3) the class membership $y_i\in\{0,1\}$ is determined by $y_i = \mathds{1}(\sset_i \geq 0.5)$. This results in a perfect AUC of $1$. (4) The perfect ordering of the class values with respect to individual scores is broken by flipping labels randomly. A set of indexes $\mathcal{I}$ of size $\lfloor \gamma n \rfloor$ is selected for which the corresponding labels are replaced by $y_i \sim \text{Ber}(0.5)$, $\forall i \in \mathcal{I}$. The fraction $\gamma$ is sampled from a $U[0.5;1]$ distribution. (5) For comparison, the empirical AUC is calculated from the vector of scores $\sset$ and flipped labels $y$. (6) The non-disclosing distributed process described in Section~\ref{subsec:gen-princ} is used to calculate the $\AUCdist$ and $\widetilde{ci}_{0.05}$. The examined values for the distributed ROC-GLM are described in Section~\ref{subsubsec:results-AUC}. The simulation is repeated $10000$ times.

\begin{figure}[!h]
    \centering
    \includegraphics[width=0.95\textwidth]{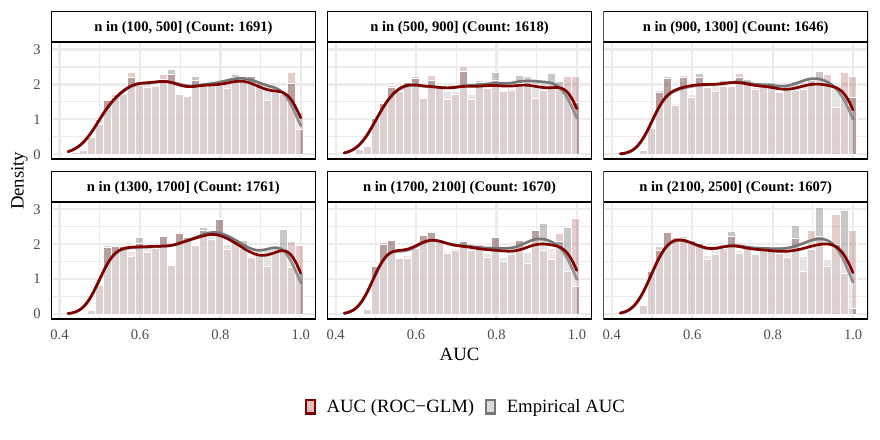}
    \caption{Densities of 10~000 simulated values of the empirical AUC and AUC from the ROC-GLM. The densities are grouped by different data sizes $n$.}
    \label{fig:auc-emp-distribution}
\end{figure}

To demonstrate the effectiveness of the basic non-distributed ROC-GLM AUC estimation, Figure~\ref{fig:auc-emp-distribution} shows the empirical distribution of the empirical as well as ROC-GLM-based AUC values depending on the sizes of $n$. The distribution of the empirical AUC values is close to the uniform distribution over the range of 0.5 to 1. The behaviour of the distribution at the borders can be explained as follows: To obtain an AUC value of one, it is necessary to keep all original class labels $y$. However, this happens rarely, due to the randomized assignment of the observations chosen in $\mathcal{I}$. The same applies to AUC values close to $0.5$. An AUC value of 0.5 appears if the class labels are completely randomized. This is also a rare event.

\subsection{Results}

\subsubsection{\rsoneText}\label{subsubsec:results-AUC}

\paragraph{ROC-GLM}

Figure~\ref{fig:auc-emp-distribution} shows a nearly perfect overlap of the empirical distributions of the empirical as well as basic non-distributed ROC-GLM-based AUC values in the range of values between 0.6 and 0.8. Nevertheless, the behaviour at the right border results from the fact that the response $U$ of the probit regression contains only very few values of zero and mostly values of 1, resulting in an unbalanced data situation. This impairs the numerical behaviour of the probit regression estimation.

Next, we quantify the difference between the empirical and the basic non-distributed ROC-GLM-based AUC estimates: $(\AUC - \AUC_{\text{ROC-GLM}})$. Table~\ref{tab:cat-emp-auc} shows summary statistics of these differences organized by bins of the empirical AUC of width 0.025. In \rsone, an absolute difference below 0.01 is requested, which is fulfilled over the whole AUC range. The mean and median differences for AUC values ranging from $0.5$ to $0.95$ fulfil this requirement, whereas AUC values between $0.95$ and $0.975$ show slightly larger differences.

\begin{table}[ht]
    \centering\footnotesize
\begin{tabular}{l|r|r|r|r|r|r|r|r}
\hline\hline
\textbf{Emp. AUC (Bin)} & \textbf{Min.} & \textbf{1st Qu.} & \textbf{Median} & \textbf{Mean} & \textbf{3rd Qu.} & \textbf{Max.} & \textbf{Sd.} & \textbf{Count}\\
\hline\hline
$(0.5,0.525]$ & $-0.0044$ & $-0.0002$ & $0.0002$ & $0.0003$ & $0.0008$ & $0.0053$ & $0.0009$ & $384$\\
\hline
$(0.525,0.55]$ & $-0.0052$ & $0.0000$ & $0.0006$ & $0.0006$ & $0.0011$ & $0.0042$ & $0.0010$ & $490$\\
\hline
$(0.55,0.575]$ & $-0.0031$ & $0.0003$ & $0.0009$ & $0.0009$ & $0.0015$ & $0.0052$ & $0.0010$ & $463$\\
\hline
$(0.575,0.6]$ & $-0.0018$ & $0.0006$ & $0.0012$ & $0.0012$ & $0.0017$ & $0.0052$ & $0.0010$ & $481$\\
\hline
$(0.6,0.625]$ & $-0.0044$ & $0.0009$ & $0.0015$ & $0.0014$ & $0.0020$ & $0.0064$ & $0.0010$ & $485$\\
\hline
$(0.625,0.65]$ & $-0.0039$ & $0.0012$ & $0.0017$ & $0.0017$ & $0.0022$ & $0.0069$ & $0.0010$ & $501$\\
\hline
$(0.65,0.675]$ & $-0.0031$ & $0.0013$ & $0.0018$ & $0.0018$ & $0.0023$ & $0.0068$ & $0.0011$ & $503$\\
\hline
$(0.675,0.7]$ & $-0.0022$ & $0.0012$ & $0.0018$ & $0.0018$ & $0.0023$ & $0.0064$ & $0.0010$ & $465$\\
\hline
$(0.7,0.725]$ & $-0.0082$ & $0.0010$ & $0.0016$ & $0.0016$ & $0.0023$ & $0.0070$ & $0.0012$ & $523$\\
\hline
$(0.725,0.75]$ & $-0.0031$ & $0.0008$ & $0.0015$ & $0.0014$ & $0.0021$ & $0.0087$ & $0.0012$ & $485$\\
\hline
$(0.75,0.775]$ & $-0.0058$ & $0.0004$ & $0.0011$ & $0.0010$ & $0.0018$ & $0.0053$ & $0.0013$ & $501$\\
\hline
$(0.775,0.8]$ & $-0.0053$ & $-0.0003$ & $0.0004$ & $0.0005$ & $0.0012$ & $0.0088$ & $0.0015$ & $523$\\
\hline
$(0.8,0.825]$ & $-0.0061$ & $-0.0013$ & $-0.0002$ & $-0.0004$ & $0.0005$ & $0.0045$ & $0.0016$ & $476$\\
\hline
$(0.825,0.85]$ & $-0.0125$ & $-0.0023$ & $-0.0013$ & $-0.0014$ & $-0.0003$ & $0.0059$ & $0.0019$ & $484$\\
\hline
$(0.85,0.875]$ & $-0.0111$ & $-0.0037$ & $-0.0026$ & $-0.0025$ & $-0.0014$ & $0.0074$ & $0.0020$ & $520$\\
\hline
$(0.875,0.9]$ & $-0.0136$ & $-0.0056$ & $-0.0044$ & $-0.0043$ & $-0.0030$ & $0.0076$ & $0.0023$ & $534$\\
\hline
$(0.9,0.925]$ & $-0.0195$ & $-0.0080$ & $-0.0065$ & $-0.0065$ & $-0.0052$ & $0.0066$ & $0.0026$ & $515$\\
\hline
$(0.925,0.95]$ & $-0.0193$ & $-0.0105$ & $-0.0091$ & $-0.0089$ & $-0.0076$ & $0.0056$ & $0.0030$ & $481$\\
\hline
$(0.95,0.975]$ & $-0.0227$ & $-0.0138$ & $\bm{-0.0113}$ & $\bm{-0.0113}$ & $-0.0093$ & $0.0067$ & $0.0037$ & $503$\\
\hline
$(0.975,1]$ & $-0.0180$ & $-0.0093$ & $-0.0062$ & $-0.0064$ & $-0.0034$ & $0.0013$ & $0.0039$ & $529$\\
\hline\hline
\end{tabular}\normalsize

    \caption{Minimum, 0.25-quantile/1st quantile, median, mean, 0.75-quantile/3rd quantile, maximum, standard deviation, and the differences $\AUC - \AUC_{\text{ROC-GLM}}$ of the bins containing the respective subset of the $10000$ empirical AUC values. Bold values indicate that these AUC bins are not smaller than 0.01 as demanded by \rsone. The count column indicates the number of simulated AUC values per bin.}
    \label{tab:cat-emp-auc}
\end{table}

\paragraph{Distributed ROC-GLM}

In the following, we investigate the accuracy of the AUC estimated by the distributed ROC-GLM. Differential privacy -- a necessary component -- is determined by the parameters $\varepsilon$ and $\delta$. These parameters must be determined in such a way that \rsone holds. The data are distributed over five \sites: The simulated prediction scores $\mathcal{F}$ and true classes $y$ are randomly split into $K = 5$ parts $\mathcal{F}^{(1)}, \dots, \mathcal{F}^{(5)}$ and $y^{(1)}, \dots, y^{(5)}$. Our simulation setting uses $\varepsilon\in A_\varepsilon = \{0.1, 0.2, 0.3, 0.4, 0.5\}$ and $\delta\in A_\delta = \{0.1, 0.2, 0.3, 0.4, 0.5\}$. Due to the Gaussian mechanism, we must also take the $\ell_2$-sensitivity into account. We assume $\ltwosens\in A_{\ltwosens} = \{0.01, 0.03, 0.05, 0.07, 0.09\}$. For the simulation, each setting of the grid $A_\varepsilon \times A_\delta \times A_{\ltwosens}$ is evaluated by simulating $10000$ data sets (cf.~Section~\ref{subsec:simulation-generation}) and hence obtaining $10000$ $\AUCdist$ values that are compared to the respective empirical AUC.

\begin{figure}[ht]
    \centering
    \includegraphics[width=0.95\textwidth]{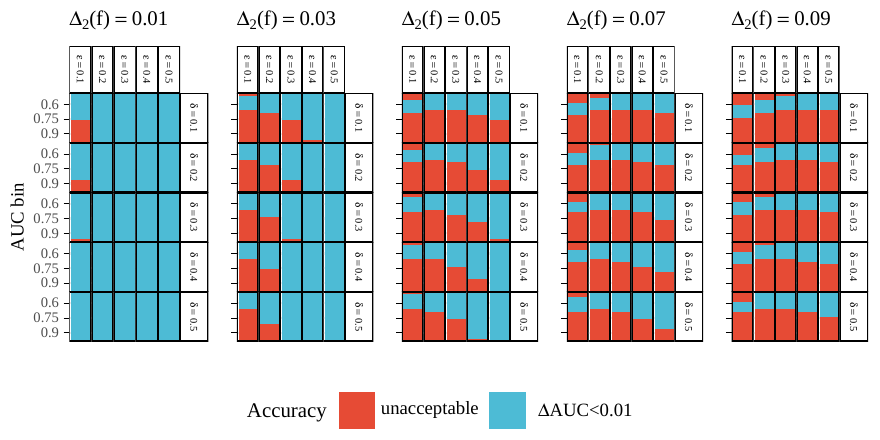}
    \caption{Combinations of the privacy parameters $\varepsilon$ and $\delta$ and their applicability. Each rectangle contains empirical AUC bins of size $0.025$ (cf.~Table~\ref{tab:cat-emp-auc}) and visualizes the mean of the absolute difference $|\Delta AUC|$ (mean absolute error, MAE) of the distributed AUC compared to the empirical AUC per bin. Each rectangle corresponds to one simulation setting $(\ltwosens, \varepsilon, \delta)$. The MAE per bin is categorized according to our hypothesis, with blue visualizing an $\text{MAE} \leq 0.01$ (\rsone) while red shows an unacceptable accuracy measured as MAE larger than $0.01$.}
    \label{fig:aucs-distr}
\end{figure}

Figure~\ref{fig:aucs-distr} shows the simulation results for different $\varepsilon$ and $\delta$ combinations. The absolute difference of the empirical AUC on the pooled data and the AUC based on the distributed ROC-GLM is checked for having a value below 0.01. The results are based on $10000$ simulation runs for 25 $\varepsilon-\delta$-combinations and for each $\ltwosens\in\{0.01, 0.03 0.05, 0.07, 0.09\}$.

Figure~\ref{fig:aucs-distr} reveals that the bias between empirical and distributed AUC depends on the $\ell_2$-sensitivity. The smaller the sensitivity and hence the better the model $\hat{f}$, less noise is required to ensure privacy. Correspondingly, smaller choices of privacy parameters can and should be used to ensure privacy. Based on the results, we choose $(\varepsilon, \delta) = (0.2, 0.1)$ for $\ltwosens \leq 0.01$, $(\varepsilon, \delta) = (0.3, 0.4)$ for $\ltwosens\in (0.01, 0.03]$, $(\varepsilon, \delta) = (0.5, 0.3)$ for $\ltwosens\in (0.03, 0.05]$, and $(\varepsilon, \delta) = (0.5, 0.5)$ for $\ltwosens\in (0.05, 0.07]$. Based on the simulation, we recommend using our distributed approach for settings with $\ltwosens > 0.07$ with caution, and we highlight that the accuracy of the AUC estimator suffers because of too much generated noise.

\subsubsection{\rstwoText}\label{subsub:corr-cis}

\begin{figure}[ht]
    \centering
    \includegraphics[width=0.95\textwidth]{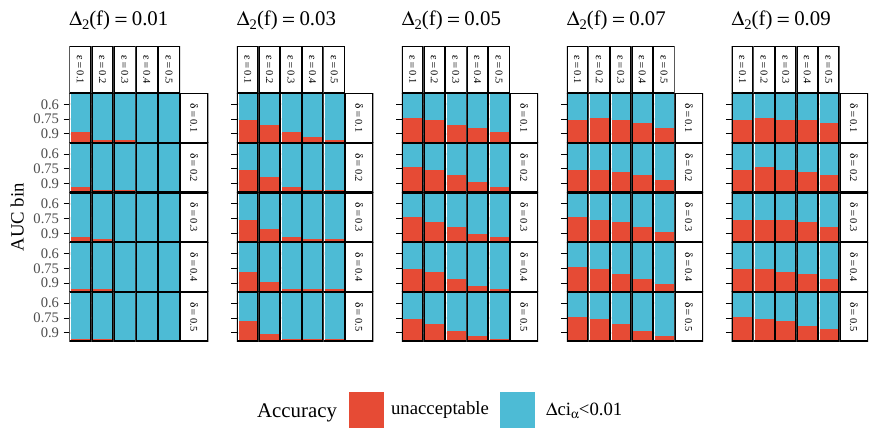}
    \caption{Combinations of the privacy parameters $\varepsilon$ and $\delta$ and their applicability depending on $\ltwosens$. Each rectangle contains empirical AUC bins of size $0.025$ (cf.~Table~\ref{tab:cat-emp-auc}) and visualizes the mean of the relative error $\Delta \ci_{0.05}$ of the distributed CI $\widetilde{ci}_{0.05}$ compared to $\ci_{0.05}$. Blue shows accuracy values with $\Delta \ci_{0.05} \leq 0.01$ (\rstwo applies), while red visualizes inaccuracies of $\Delta \ci > 0.01$.}
    \label{fig:cis-distr}
\end{figure}

The respective results in terms of acceptable $(\varepsilon, \delta)$ combinations are shown in Figure~\ref{fig:cis-distr}. Acceptable $(\varepsilon, \delta)$ combinations under \rsone are also acceptable under \rstwo. Therefore, we recommend using the more restrictive settings described in the previous Section~\ref{subsubsec:results-AUC} for the AUC estimation of the distributed ROC-GLM.

\section{Data analysis}\label{sec:use-case}

\newcommand{\cidistlower}{0.6051}
\newcommand{\cidistupper}{0.7595}
\newcommand{\ciemplower}{0.6131}
\newcommand{\ciempupper}{0.7608}
\newcommand{\aucdist}{0.6875}
\newcommand{\aucpooled}{0.6919}
\newcommand{\rocglmparamOne}{0.7817}
\newcommand{\rocglmparamTwo}{1.2486}
\newcommand{\bsemp}{0.1733}
\newcommand{\ts}{730}
\newcommand{\nOne}{56}
\newcommand{\nTwo}{49}
\newcommand{\nThree}{60}
\newcommand{\nFour}{49}
\newcommand{\nFive}{60}
\newcommand{\privparOne}{0.3}
\newcommand{\privparTwo}{0.4}
\newcommand{\ltwosensUC}{0.016}
\newcommand{\AUCdiffusecase}{0.0044}
\newcommand{\CIdiffusecase}{0.0094}

In this chapter, we develop a prognostic model and validate its predictive performance on a distributed test data set. The following presents the distributed analysis, which is also compared to the pooled analysis (see Section~\ref{subsec:analysis-pooled}). As a privacy level, we choose a value of $q = 5$ (see Section~\ref{subsec:gen-princ}, \Aone).

\paragraph{About the data}

The data set is provided by the German Breast Cancer Study Group~\citep{schumacher1994randomized} and can be found in the \texttt{TH.data} package~\citep{TH.data}. The data consists of records from 686 breast cancer patients on the  effect of hormonal therapy on survival. Besides the binary variable hormonal treatment (\texttt{horTH}), the data set provides information on age (\texttt{age}), menopausal status (\texttt{menostat}), tumor size (in mm, \texttt{tsize}), tumor grade  (\texttt{tgrade}), number of positive nodes (\texttt{pnodes}), progesterone receptor (in fmol, \texttt{progrec}), estrogen receptor (in fmol, \texttt{estrec}), recurrence-free survival time (in days, \texttt{time}), and censoring indicator (0- censored, 1- event, \texttt{cens}).

Because the data set is (by its nature) not distributed, we use 60 \% ($412$ observations) for training the model and split the remaining 40 \% ($274$ observations) into 5 parts $\datak{1}, \dots, \datak{5}$ with $\nk[1] = \nOne$, $\nk[2] = \nTwo$, $\nk[3] = \nThree$, $\nk[4] = \nFour$, and $\nk[5] = \nFive$ that are used for the distributed validation. Each split is distributed to a \site to simulate the distributed setup.

The aim is to predict the survival probability $p(t|\bm{x}) = P(T > t| X = \bm{x})$ of surviving time point $t$ based on covariates $\bm{x}$. For the use case, we choose $t = \ts$ (two years), and therefore, the goal is to validate the survival probability of a patient after two years in the study. The predicted scores are the survival probabilities $\hat{y_i} = \fh(\bm{x}_i) = \hat{p}(\ts|\bm{x}_i)$ with $\bm{x}_i\in\cup_{k=1}^K \datak{k}$. The corresponding binary variable $y_i$ equals $0$ if the patient dies in $[0, \ts]$ or a recurrence was observed, and $y_i$ equals $1$ if otherwise. Therefore, a high value for the survival probability $\hat{y}_i$ ideally corresponds to a binary outcome of $1$.

\paragraph{About the model}

We choose a random forest~\citep{breiman2001random} using the \texttt{R} package \texttt{ranger}~\citep{ranger} as a prognostic model $\fh$ for the survival probability $p(t|\bm{x})$. With the exception of the number of trees (which is set to $20$), the random forest was trained with the default hyperparameter settings of the \texttt{ranger} implementation. The model formula is given by
$$
\textsf{Surv(time, cens)} \sim \textsf{horTh + age + tsize + tgrade + pnodes + progrec + estrec}.
$$

\paragraph{About the implementation}

The implementation is based on the DataSHIELD~\citep{gaye2014datashield} framework and is provided by an \texttt{R} package called \texttt{dsBinVal} (\url{github.com/difuture-lmu/dsBinVal}). Further details about these methods and privacy considerations can be found in the respective GitHub README.

\paragraph{Aim of the analysis}

The main goal of the analysis is to test the hypothesis that the true AUC is significantly larger than $0.6$ as the minimal prognostic performance of the model $\fh$. The significance level is set to $\alpha = 0.05$:
\begin{equation}\label{eq:app-hypothesis}
    H_0: \ \AUC \leq 0.6 \ \ \text{vs.} \ \ H_1: \ \AUC > 0.6
\end{equation}

To test the hypothesis, we estimate the AUC with $\AUCdist$ using the distributed ROC-GLM as well as the approximated CI $\widetilde{\ci}_{0.05}$. We reject $H_0$ if $\AUC > 0.6,\ \forall \AUC \in \widetilde{\ci}_{0.05}$.

\paragraph{Analysis plan}

In the following, (1) we start in with the calculation of the $\ell_2$-sensitivity (Section~\ref{subsec:app-privacy-choice}). Depending on the result, we set the privacy parameters $\varepsilon$ and $\delta$. Next, (2) we continue with fitting the distributed ROC-GLM and calculating the approximation of the AUC CI (Section~\ref{subsec:app-distr-roc-glm}). At this point, we are able to make a decision about the hypothesis in equation~\eqref{eq:app-hypothesis}. In a final step, (3) we demonstrate how to check the calibration of the model using the distributed Brier score and calibration curve (Section~\ref{subsec:app-calibration}).

\subsection{Choice of the privacy parameters}\label{subsec:app-privacy-choice}

Given the model and the data set, the $\ell_2$-sensitivity is $\ltwosens = \ltwosensUC$. Following the results of Section~\ref{subsubsec:results-AUC}, we use $\varepsilon = \privparOne$ and $\delta = \privparTwo$, as suggested for $\ltwosens \in (0.01, 0.03]$.

\subsection{Calculation of the distributed ROC-GLM}\label{subsec:app-distr-roc-glm}

\begin{figure}[ht]
    \centering
    \includegraphics{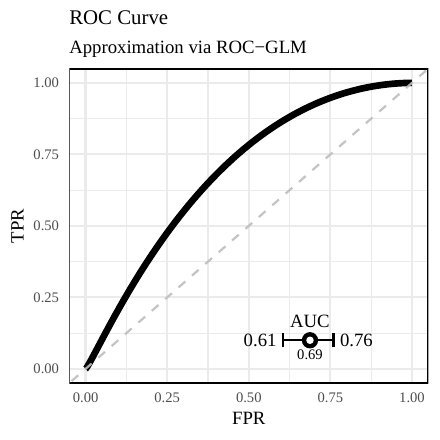}
    \caption{ROC curve estimated by the distributed ROC-GLM.}
    \label{fig:analysis-distr-roc-glm}
\end{figure}

The fit of the ROC-GLM results in parameter estimates of $\gamma_1 = \rocglmparamOne$ and $\gamma_2 = \rocglmparamTwo$. The AUC obtained from the ROC curve using these parameters is $\AUC_{\text{ROC-GLM}} = \aucdist$ with $\widetilde{\ci}_{0.05} = [\cidistlower, \cidistupper]$. The results are visualized in Figure~\ref{fig:analysis-distr-roc-glm}.

Based on the given CI, we significantly reject $H_0$ for $H_1$ and hence assume the true AUC to be greater than 0.6.

\subsection{Checking the calibration}\label{subsec:app-calibration}

\begin{figure}[ht]
    \centering
    \includegraphics{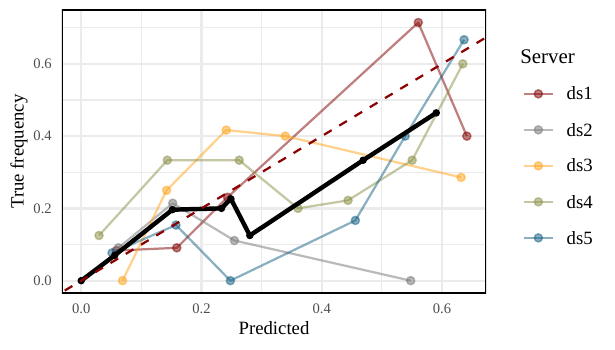}
    \caption{Calibration curve (bold line) and calibration curves of the individual \sites using $10$ bins. Note that aggregated values from the \site are only shared if one bin contains more than $5$ values. See Appendix~A.3 for tables containing the numbers of values per bin.}
    \label{fig:analysis-calibration}
\end{figure}

The Brier score of $\fh$ calculates to $\operatorname{BS} = \bsemp$ and indicates a good but not perfect calibration. We further assume our model to be not calibrated perfectly. Still, the calibration is adequate, but the model seems to underestimate the true relative frequencies for scores greater than 0.3. Figure~\ref{fig:analysis-calibration} shows the distributed calibration curve as well as the individual calibration curves per \site. Furthermore, we observe that the range of the calibration curve does not cover the whole range of the scores $\fh(x) \in [0, 1]$. This indicates that our model does not predict scores close to $1$. We want to highlight that, due to privacy reasons, not all score values were included in the calculation; aggregated values are only shared if they consist of at least $5$ elements. The table in Appendix~A.3 shows the number of elements per bin and \site.

\subsection{Comparison with pooled data}\label{subsec:analysis-pooled}

\begin{figure}[ht]
\centering
\begin{minipage}{.5\textwidth}
  \centering
  \includegraphics[width=.8\linewidth]{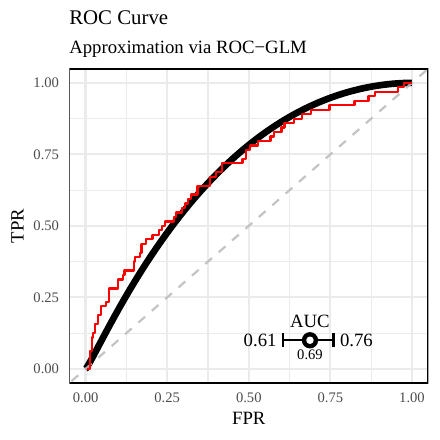}
\end{minipage}%
\begin{minipage}{.5\textwidth}
  \centering
  \includegraphics[width=.8\linewidth]{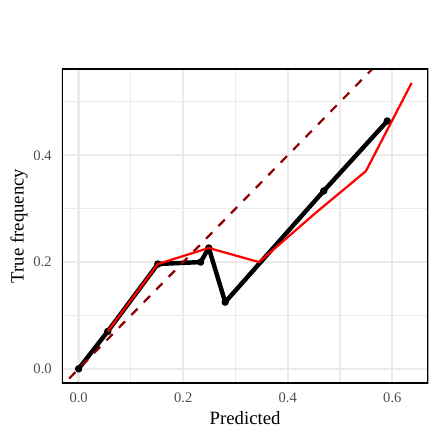}
\end{minipage}
\caption{Comparison of the empirical ROC curve with ROC curve obtained by the distributed ROC-GLM (left). Comparison of the calibration curve when calculated on the pooled scores compared with the distributed calibration curve (right). The thin curves are the lines on the pooled data.}
\label{fig:pooled}
\end{figure}

Comparing the ROC curves using the empirical ROC and the distributed ROC-GLM (Figure~\ref{fig:pooled}, left) shows a good fit of the ROC-GLM. The resulting AUC values are $\AUCdist = \aucdist$ and $\AUC = \aucpooled$ with $|\Delta AUC| = \AUCdiffusecase < 0.01$. The CIs of the approximated CI $\widetilde{\ci}_{0.05} = [\cidistlower, \cidistupper]$ and the CI on the pooled scores $\ci_{0.05} = [\ciemplower, \ciempupper]$ reveals a slightly more pessimistic CI estimation in the distributed setup. The error of the CI calculates to $\Delta \ci_{0.05} = \CIdiffusecase < 0.01$.

The distributed calibration curve shows a good overlap with the calibration curve in areas where all data are allowed to be shared. For bins where this is not the case, the distributed calibration curve is off. Still, the tendency of over- or underestimation of the distributed calibration curve corresponds to one of the pooled curves. The bins for which the full information was received are $[0, 0.1]$, $(0.1, 0.2]$, and $(0.2, 0.3]$ (cf. Appendix~A.3 table~1). For all other bins, at least one \site was not allowed to share the aggregated values. The Brier score of the pooled and distributed approach is equal.

\section{Reproducibility considerations}

All experiments were conducted using \texttt{R} version 4.1.2 on a Linux machine with an Intel(R) Core(TM) i7-8665U CPU @ 1.90GHz processor. The package used to run the simulation was \texttt{batchtools}~\citep{batchtools}. The code to reproduce all results as well as all simulation results is available in a GitHub repository\footnote{\url{github.com/difuture-lmu/simulations-distr-auc}}. The repository contains a README file with further details and a script to install all packages with the respective version used when the benchmark was conducted. Furthermore, a Docker image\footnote{\url{hub.docker.com/repository/docker/schalkdaniel/simulations-distr-auc}} can be installed providing a snapshot of the system at the time of the benchmark containing \texttt{R} and all packages with their respective version. The Docker image also comes with the RStudio container\footnote{\url{hub.docker.com/r/rocker/rstudio}} that allows direct inspection of all the results in a web browser.

The code to conduct the data analysis is given in a separate GitHub repository\footnote{\url{github.com/difuture-lmu/datashield-roc-glm-demo}}. The repository contains the data, an installation of all necessary packages, as well as code to set up the publicly available DataSHIELD server\footnote{\footnotesize{Available at \url{opal-demo.obiba.org}. The reference, username, and password are available at the OPAL documentation \url{opaldoc.obiba.org/en/latest/resources.html} in the \enquote{Types} section.}} to run the analysis\footnote{\footnotesize{We cannot guarantee the functionality of the DataSHIELD server or if it will be publicly available forever. However, we keep the repository up-to-date by using continuous integration, which is triggered automatically every week. This system also reports errors that occur if the analysis cannot be conducted on the test server anymore. Further information can be found in the README file of the repository.}}.

\section{Discussion}

Distributed non-disclosing (i.e., privacy-preserving) strategies for data analysis are highly relevant for data-driven biomedical research. Since the analyses can be considered anonymous, current legal data protection frameworks allow their use without requesting specific consent. Protecting privacy by appropriate means is fundamental when using personal data for research. These technologies also enable taking part in broader network structures without additional administrative work concerning data protection issues. Privacy-preserving distributed computation allows researchers to digitally cooperate and leverage the value of their data while respecting data sovereignty and without compromising privacy. Besides the privacy preservation in algorithms that are backed up with security mechanisms, it is worth noting that software is also a key player in privacy-preserving analysis. For example, most models fitted with the statistical software \texttt{R} attach data directly to the model object. Sharing these objects without caution gives analysts direct access to the training data~\citep[cf., e.g.,][]{ibc2022schalk}.

International activity has been dedicated to setting up distributed non-disclosing analysis frameworks, which implement machine learning approaches into a distributed analysis scheme. The availability of the respective algorithms is growing, and distributed learning for data from heterogeneous clinical servers has emerged as a hot field. However, our impression is that algorithms for distributed validation of these learning algorithms are lacking.

In this paper, we specifically focused on the assessment of discrimination and calibration of learning algorithms with a binary outcome. The discrimination is estimated by a ROC curve and its AUC. We also provide CIs to the distributed AUC estimate. The distributed estimation process is based on \textit{placement values} and \textit{survivor functions}. They represent qualities of the global distribution of score values (aggregated over all centers). To do this in a non-disclosing way, we applied differential privacy techniques. With the creation of the placement values and the transmission of this information to the local server, we applied a distributed version of the ROC-GLM approach to estimate the ROC curve and its AUC in a distributed way. We used a straightforward approach for the distributed GLM estimation. However, we acknowledge that there may be more efficient approaches, and we will explore this aspect in future work.

\noindent {\bf{Abbreviations}}

\noindent {\it{AUC: Area under the curve; CI: Confidence interval; DP: Differential privacy; FPR: False positive rate; GLM: Generalized linear model; IPD: Individual patient data; MII: Medical Informatics Initiative; ROC: Receiver operating characteristics; TPR: True positive rate.}

\section*{Declarations}

\noindent {\bf{Ethical Approval and consent to participate}}

\noindent {\it{Not applicable.}}

\noindent {\bf{Consent to Publication}}

\noindent {\it{Not applicable.}}

\noindent {\bf{Data Availability statement}}

\noindent {\it{The simulated datasets generated during the current study are available on GitHub, \url{https://github.com/difuture-lmu/simulations-distr-auc}.}}

\noindent {\bf{Conflict of interest}}

\noindent {\it{The authors declare no competing interests.}}

\noindent {\bf{Funding}}

\noindent {\it{Not applicable.}}

\noindent {\bf{Acknowledgment}}

\noindent {\it{This work was supported by the German Federal Ministry of Education and Research (BMBF) under Grant No. 01IS18036A and Federal Ministry for Research and Technology (BMFT) under Grant No. 01ZZ1804C (DIFUTURE, MII). The authors of this work take full responsibility for its content.}}

\noindent {\bf{Author contribution}}

\noindent {\it{DS wrote the manuscript, implemented the methods and the simulation study, and prepared the use case. DS also created all graphics and the interpretation of the results from the simulation study and from the use case. In addition, much effort was put into reproducibility, for which DS created a Docker container and a GitHub repository with the simulation study results. The idea of the distributed AUC calculation originated from UM. All co-authors provided substantial assistance in writing the manuscript and interpreting the simulation study results.}}

\bibliographystyle{apalike}
\bibliography{references}

\newpage
\phantom{aaaa}
\end{document}